\documentclass[12pt]{iopart}
\usepackage{iopams,bm,verbatim}
\begin{document}

\title{Complete integration of the aligned Newman Tamburino Maxwell solutions}
\author{L.~De Groote and N.~Van den Bergh }

\address{Department of Mathematical Analysis IW16, Ghent University, Galglaan 2, 9000 Ghent, Belgium}

\begin{abstract}
We investigate the cylindrical class of Newman Tamburino equations in the presence of an aligned Maxwell field. After obtaining a complete integration of the field equations we look at the possible vacuum limits and we examine the symmetries of the general solution.
\end{abstract}

\section{Introduction}
In a previous work we examined the generalization of the Newman Tamburino metrics in the presence of an aligned Maxwell field~\cite{DeGrooteVdB}. It was shown there that the so called `spherical' class did not admit any solutions and that consistency of the field equations therefore required the `cylindrical' condition $|\rho|=|\sigma|$. The existence of solutions in the cylindrical class was left however as an open question.

\par
In the present paper we obtain a complete integration of the Einstein Maxwell field equations for this problem. We first show that the Newman-Penrose, Bianchi and Maxwell equations form an integrable system and then proceed to integrate the first Cartan structure equations.  After obtaining the general solution we look at the limiting case, where the charge Q of the Maxwell field goes to zero. Surprisingly in this limit, we do not recover the empty space metric obtained by Newman and Tamburino, but rather the special case admitting two Killing vectors.

\par
In \S 2 we describe the choice of tetrad and the integration method. The resulting metric is presented at the end of this section. In \S 3 we discuss the vacuum limit of the metric and its symmetry properties.

\section{Cylindrical class}

We follow the notations and conventions of our previous paper~\cite{DeGrooteVdB}.
The cylindrical class of Newman Tamburino metrics is characterized by the existence of a principal null direction $\mathbf{k}$ of the Weyl tensor which is hypersurface orthogonal and geodesic,
\begin{equation}
\Psi_0 = 0, \label{psi0}
\end{equation}
\begin{equation}
\kappa = 0, \label{kappanul}
\end{equation}
\begin{equation}
\rho - \bar{\rho} = 0, \label{rho}
\end{equation}
but has nonvanishing shear and divergence, with the spin coefficients $\rho$ and $\sigma$ related by
\begin{equation}
\rho^2 - |\sigma|^2 = 0. \label{cyl}
\end{equation}
We assume now that a non-null Maxwell field is present, which is aligned in the sense that $\mathbf{k}$ is also a principal null vector of the Maxwell tensor :
\begin{equation}
\Phi_0 = 0. \label{phi0}
\end{equation}
As mentioned in~\cite{DeGrooteVdB} we can, by means of a rotation, a boost and a null rotation, set up the complex null tetrad $\left( \mathbf{m}, \mathbf{\bar{m}}, \mathbf{l}, \mathbf{k} \right)$ such that it is parallelly propagated along $\mathbf{k}$ :
\begin{equation}
\kappa = \epsilon = \pi = 0
\end{equation}
and such that, additionally, 
\begin{equation}\tau = \bar{\alpha} + \beta.
\end{equation}

The NP equations guarantee then that $D(\sigma / \bar{\sigma} )=0$ and hence, by (\ref{cyl}), a spatial rotation exists such that
\begin{equation}
\sigma = \rho.
\end{equation}
The remaining tetrad freedom consists now of boosts and null rotations with parameters $A$ and $B$ respectively satisfying :
\begin{equation}
\label{parameters}
D\,A =  \delta\,A = \bar{\delta}\,A = 0 = D\, B.
\end{equation}

\par
This allows us to rewrite the NP, Maxwell and Bianchi equations determining the evolution along the geodesic rays as follows :
\begin{equation}
\eqalign{
D\, \rho &=2\,{\rho}^{2},\\
D\,\alpha &=\left( \alpha+\beta \right)\rho,\\
D\, \lambda &=\left( \lambda+\mu\right)\rho ,\\
D\, \mu &=\left( \lambda+\mu\right)\rho +\Psi_{{2}}+\frac{1}{12}\,R ,\\
D\, \nu &= \left( \alpha+\bar{\beta} \right) \mu +\left( \bar{\alpha} + \beta\right) \lambda +\Psi_{{3}}+ \Phi_2 \,\overline{\Phi_1} ,\\
D\, \beta &=\left( \alpha+\beta \right)\rho+\Psi_{{1}} ,\\
D\, \gamma &=\alpha\,\bar{\alpha} +2\,\alpha\,\beta+\beta\,\bar{\beta} +\Psi_{{2}}-\frac{1}{24}\,R+\Phi_1\, \overline{\Phi_1},
}
\end{equation}
while
\begin{equation}
\label{DPhi1}
{D} \, \Phi_1 =2\,\rho\,\Phi_1 ,
\end{equation}
\begin{equation}
\label{DPsi1}
D\, \Psi_{{1}} =4\,\rho\,\Psi_{{1}} ,
\end{equation}
and
\begin{equation}
\label{deltaPhi1}\delta \, \Phi_1 =2\left( \bar{\alpha} +\beta \right) \Phi_1 -\rho\,\Phi_2 ,
\end{equation}
\begin{equation}
\label{deltaPsi1}
\delta \, \Psi_{{1}} =2\left( 2\,\bar{\alpha} +3\,\beta\right) \Psi_{{1}}-\left( 3\,\Psi_{{2}}+2\,\Phi_1 \,\overline{\Phi_1}\right)\rho .
\end{equation}

\par
Next we calculate $[\delta, D]\, \Psi_1$, from which we eliminate $D\,\Psi_2$ by $(2\,B_9 - 3\, B_5)$ :
\begin{equation*}
\Psi_{{1}}\delta\,\rho +\rho\,\bar{\delta} \,\Psi_1 +2\,{\rho}^{2}\Phi_{{1}}\overline{\Phi_1} - \left( \bar{\alpha}+\beta +3\,\alpha +\bar{\beta}\right) \rho\,\Psi_1 -\frac{3}{2} \,{\Psi_{{1}}}^{2}  =0.
\label{temp1}
\end{equation*}
The last equation, together with $[\bar{\delta}, D]\left( \rho\,\Psi_1\right) $, gives us expressions for $\delta\,\rho$ and $\bar{\delta}\,\Psi_1$ :
\begin{eqnarray}
\label{delrho} \delta\,\rho &=\frac{1}{7} \left( 5\,{\frac {\overline{\Psi_2}}{\overline{\Psi_1} }}-2 \,{\frac {\Psi_{{2}}}{\Psi_{{1}}}}\right)\rho^2 +\frac{1}{7} \left( 9 \left( \bar{\alpha} +\beta\right) - 5\left( \alpha+\bar{\beta} \right) \right) \rho
 +\frac{1}{2} \,\overline{\Psi_1},\\
\label{delbarPsi1} \bar{\delta} \,\Psi_1 &= \left( {\frac {2}{7}}\left(13\,\alpha - \beta -\bar{\alpha} +6\,\bar{\beta} \right) -\frac{5}{7} \,{\frac {\rho\,\overline{\Psi_2} }{\overline{\Psi_1} }}+\frac{1}{2} \,{\frac {3\,\Psi_1-\overline{\Psi_1} }{\rho}}\right) \Psi_1 \nonumber \\
&+\frac{2}{7} \,\rho\,\Psi_{{2}} -2\,\rho\,\Phi_1 \,\overline{\Phi_1} .
\end{eqnarray}

From $(2\,B_9 - 3\, B_5)$ we find the evolution of $\Psi_2$ along the geodesic rays :
\begin{eqnarray}
\nonumber
D\,\Psi_2 =\left( {\frac {2}{7}} \left( 6\,\alpha-\beta-\bar{\alpha} +6\,\bar{\beta}  \right) -\frac{5}{7} \,{\frac {\rho\,\overline{\Psi_2} }{\overline{\Psi_1} }}+\frac{1}{2} \,{\frac {3\,\Psi_1-\overline{\Psi_1} }{\rho}}\right) \Psi_1 +{\frac {23}{7}}\,\rho\,\Psi_{{2}}.
\end{eqnarray}
We can now simplify $NP_{11}$ and its complex conjugate by means of $[\delta, D]\,\rho$, and eliminate $\Psi_2$ from both. This results in
\begin{equation}
\Psi_{{2}}=\frac{1}{2} \,{\frac {\Psi_{{1}} \left( \alpha+3\,\beta+\bar{\alpha}-\bar{\beta} \right) }{\rho}}
\end{equation}
and
\begin{equation}\label{eq1}
2\left(\alpha -\beta + \bar{\alpha} -\bar{\beta}\right) \rho +\Psi_{{1}}+\overline{\Psi_1} =0.
\end{equation}

\par
From $[\delta, D]\,\Phi_1$ and the second Maxwell's equation we get expressions for $D\,\Phi_2$ and $\bar{\delta}\,\Phi_1$, which we can use to solve $[\bar{\delta}, D]\,\Phi_1$ for $\Phi_2$ :
\begin{equation}
\eqalign{
D\,\Phi_2 &=\frac{1}{2} \left( 3\,\alpha + \beta -\bar{\alpha} +\bar{\beta} +{\frac {2\,\Psi_{{1}} - \overline{\Psi_1}}{\rho}} \right) \Phi_1  +\rho\,\Phi_2 ,\\
\bar{\delta}\,\Phi_1 &=\frac{1}{2} \left( 3\,\alpha + \beta -\bar{\alpha} +\bar{\beta} +{\frac {2 \,\Psi_{{1}}- \overline{\Psi_1}}{\rho}} \right) \Phi_1,\\
\Phi_2 &=\frac{1}{2} \left( \alpha+ 3\,\beta+\bar{\alpha} -\bar{\beta}- \frac{\Psi_1}{\rho}\right)\frac{\Phi_1}{\rho}.
}
\end{equation}

\par
In the next step we use (\ref{deltaPsi1},\, \ref{delrho},\, \ref{delbarPsi1}) for evaluating $[\bar{\delta},\, \delta] \Psi_1$ and $[\bar{\delta},\, \delta] \rho$. Herewith we obtain
\begin{equation}\label{stapje1}
\eqalign{
\delta\,(\beta +\bar{\beta}) &=\delta\,\bar{\alpha}+ 2\,\bar{\delta}\,\bar{\alpha} -\delta\,\alpha  +\left( \alpha -{\beta}+ 3\,\bar{\beta} -4\,{\frac {\rho\,\Phi_1 \,\overline{\Phi_1}  }{\Psi_{{1}}}} \right) \beta \\
&- \left( 2\,\alpha +3\,\bar{\alpha}-4\, \bar{\beta} +2\,{\frac {\Psi_{{1}}}{\rho}}\right) \bar{\beta} +2\left( \mu - \bar{\mu} + 2\,{\frac {\alpha\,\Phi_1 \,\overline{\Phi_1}  }{\Psi_{{1}}}} \right) \rho \\
&-2\,{\alpha}^{2} -\alpha\,\bar{\alpha} + \bar{\alpha}^{2} -\left( \alpha -\bar{\alpha} \right) {\frac {\Psi_{{1}}}{\rho}}
}
\end{equation}
and
\begin{equation}\label{stapje2}
\eqalign{
\bar{\delta}(\beta +\bar{\beta}) &=\bar{\delta}\,{\alpha} - \bar{\delta}\,\bar{\alpha} +2\delta\,\alpha  +\left( 7\alpha +2\bar{\alpha} -2{\beta} -4 {\frac {\rho\,\Phi_1 \,\overline{\Phi_1}  }{\Psi_{{1}}}} + 3 \frac{\Psi_1}{\rho} \right) \beta \\
&+ \left( 4 \,\alpha -\bar{\alpha} - 3\,\beta - \bar{\beta} +{\frac {\Psi_{{1}}}{\rho}}\right) \bar{\beta} -2\left(\mu - \bar{\mu} - 2\,{\frac {\alpha\,\Phi_1 \,\overline{\Phi_1}  }{\Psi_{{1}}}} \right) \rho \\
&-3\,{\alpha}^{2} -3\, \alpha\,\bar{\alpha} -2\, \alpha {\frac {\Psi_{{1}}}{\rho}}
}
\end{equation}
which allows us to find the $\delta\,$ derivative of (\ref{eq1}) as
\begin{equation}\label{stapje3}
\eqalign{
\bar{\delta}\,\bar{\alpha} -\delta\,\alpha  &= \left(\bar{\mu}- \mu\right)\rho + \frac{1}{4}\left( 3\,\alpha^2 +\beta^2 + \bar{\alpha}^2 - 9\,\bar{\beta}^2\right) +\alpha\,\bar{\alpha} - 2\,\beta\,\bar{\beta} \\
&+ \bar{\alpha}\,\bar{\beta}  + \frac{3}{2} \alpha\,\bar{\beta} - \frac{1}{2}\bar{\alpha}\,\beta + \frac{1}{8} \left( \alpha + 3\beta +\bar{\alpha} + 11 \bar{\beta} - \frac{\overline{\Psi_1}}{\rho} \right) \frac{\Psi_1}{\rho} \\
&+ \frac{1}{8} \left( \alpha -\beta + 9\bar{\alpha} -\bar{\beta} + 3\frac{\overline{\Psi_1}}{\rho} \right) \frac{\overline{\Psi_1}}{\rho} + \left( 2\left(\beta-\alpha\right) \frac{\rho}{\Psi_1} - 1 \right) \Phi_1\overline{\Phi_1}.
}
\end{equation}

\par
Now follows a crucial step, in the form of two beautiful factorizations. Expressing that the right hand sides of $\overline{(\ref{stapje1})}-(\ref{stapje2})$ and $\overline{(\ref{stapje3})}+(\ref{stapje3})$ are 0 and eliminating from both of these $\bar{\alpha}$ by means of equation (\ref{eq1}), we find that:
\begin{equation}
\label{eq2}
\left( 2\,\beta\,\rho-2\,\rho\,\alpha-\Psi_{{1}} \right)  \left( {\rho}^{2}\Phi_1 \,\overline{\Phi_1}  \overline{\Psi_1} +3\,{\rho}^{2}\Phi_1 \,\overline{\Phi_1}  \Psi_{{1}}+\overline{\Psi_1} {\Psi_{{1}}}^{2} \right) =0.
\end{equation}
and
\begin{equation}
\label{eq3}
\left( \overline{\Psi_1} -\Psi_{{1}} \right)  \left( 2\,\beta\,\rho-2\,\rho\,\alpha-\Psi_{{1}} \right)  \left( 2\,{\rho}^{2}\Phi_1 \,\overline{\Phi_1}  +\Psi_{{1}}\overline{\Psi_1}  \right) =0.
\end{equation}
It can easily be seen from the above results that the common factor $2\,\beta\,\rho-2\,\rho\,\alpha-\Psi_{{1}}$ has to be zero. If not, then by (\ref{eq3}) $\Psi_1$ would have to be real, after which the second factor of (\ref{eq2}) would factorize as $\Psi_1$ times a positive definite expression. We therefore conclude that
\begin{equation*}
\beta=\alpha+\frac{1}{2} \,{\frac {\Psi_{{1}}}{\rho}}.
\end{equation*}

\par
Next we show that the Ricci scalar (or, equivalently, the cosmological term) is zero: $R = 0$. In order to do so, we first rewrite $NP_{12}$ as
\begin{eqnarray*}
\eqalign{
\bar{\delta}  \, \alpha &=\delta\,\alpha +\left( \lambda - \mu \right) \rho + 2\left( \alpha -\bar{\alpha} \right) \alpha + \left( 2\,\alpha -\frac{1}{2}\,\bar{\alpha} +\frac{1}{4}\,\frac{\Psi_1-\overline{\Psi_1}}{\rho} \right) {\frac {\Psi_{{1}}}{\rho}} \\
&-\frac{1}{2} \,{\frac {\alpha \overline{\Psi_1} }{\rho}}-\frac{1}{24}\,R,
}
\end{eqnarray*}
after which the fourth Maxwell's equation reads
\begin{eqnarray}
\nonumber
\frac{1}{8} \left( 3\,{\Psi_{{1}}}^{2}- 2\,\overline{\Psi_1} \Psi_{{1}} - \overline{\Psi_1}^{2}\right) \frac{\Phi_1}{\rho^3} + \frac{1}{4} \left(   \left( \Psi_{{1}} - \overline{\Psi_1}\right) \bar{\alpha} +\left( 9 \,\Psi_{{1}} -5 \,\overline{\Psi_1}\right) \alpha \right) \frac{\Phi_1}{\rho^2}\\
+ 2 \left( \alpha + \delta\,\alpha -\bar{\alpha} \alpha \right) \frac{\Phi_1}{\rho}-2\,\mu\,\Phi_1 -\Delta \Phi_1 =0.
\label{temp}
\end{eqnarray}
If we calculate the propagation of this expression along the geodesic rays, eliminate the terms $D\,\Delta\,\Phi_1$ and $D\,\delta\,\alpha$ by means of the commutators $[\Delta, D]\Phi_1$ and $[\delta, D]\alpha$, and simplify the result by equation (\ref{temp}) and $NP_{16}$, we get
\begin{eqnarray}
\nonumber
\frac{1}{2}\left( 11\, \alpha +\bar{\alpha} +2\,\frac{\Psi_1}{\rho} \right) \frac{\Psi_1}{\rho} -\frac{1}{2} \left( \alpha + 11\,\bar{\alpha} + 2\, \frac{\overline{\Psi_1}}{\rho} \right) \frac{\overline{\Psi_1}}{\rho} +6\left(\alpha^2 -\,\bar{\alpha}^2\right) \\
 - 2\left(\mu-\bar{\mu}  + 2\,\gamma - 2\,\bar{\gamma} \right)\rho+ 2 \left( \delta\,\alpha - \bar{\delta}\,\bar{\alpha} \right) -\frac{1}{6}\,R =0.
\label{eq1_2}
\end{eqnarray}
Adding (\ref{eq1_2}) and its complex conjugate results in $R = 0$.

\par
From $[\delta, D]\,\alpha$, $NP_{16}$ and ($3\,B_8+2\,B_{10}$) we now see that respectively
\begin{equation*}
{D} \delta\alpha = 4\rho\delta\alpha +\left(\lambda - \mu\right) {\rho}^{2} +\left( \alpha+ \bar{\alpha} +\frac{1}{8} {\frac {3\,\Psi_1 + \overline{\Psi_1} }{\rho}}\right) \Psi_{{1}}+\alpha  \overline{\Psi_1} -\rho\,\Phi_1 \overline{\Phi_1} ,
\end{equation*}
\begin{eqnarray*}
\eqalign{
\Delta \rho &= -\frac{1}{4} \left( \bar{\alpha} +5\,\alpha + \frac{\overline{\Psi_1}}{\rho} \right) \frac{\Psi_1}{\rho} -\frac{1}{4} \left( 3\,\alpha - 5\,\bar{\alpha} \right) \frac{\overline{\Psi_1}}{\rho} +\left( 3\,\gamma - \bar{\gamma} -2\,\mu \right) \rho \\
&
+ \bar{\alpha} ^{2} - 4\,\alpha\,\bar{\alpha} - {\alpha}^{2} + 2\,\delta\,\alpha -\Phi_1\overline{\Phi_1}
}
\end{eqnarray*}
and
\begin{eqnarray*}
\eqalign{
\Delta \Psi_{{1}} &= \frac{1}{8} \left(3\,\overline{\Psi_1} -5\,\Psi_1 \right) {\frac {{\Psi_{{1}}}^{2}}{{\rho}^{3}}}+\frac{1}{4} \left( \left( 3\,\bar{\alpha} - 25\,\alpha \right) \Psi_1 + \left( 3\,\alpha + 7\,\bar{\alpha} \right) \overline{\Psi_1} \right) \frac{\Psi_1}{\rho^2}\\
& + \left( 2\,\delta\,\alpha -8\,\alpha^2 -\alpha\,\bar{\alpha} +2\,\bar{\alpha}^2 - \Phi_1\overline{\Phi_1} \right) \frac{\Psi_1}{\rho}+2 \left( 2\,\gamma -\bar{\gamma} -\mu \right) \Psi_1 \\
& +2\,\rho\,\Psi_{{3}} -2\left( \alpha + \bar{\alpha}\right) \Phi_1 \overline{\Phi_1}.
}
\end{eqnarray*}
Then we rewrite the Bianchi identity ($3\,B_7+2\,B_{10}$)
as an equation for $D\,\Psi_3$ :
\begin{eqnarray}
\eqalign{
D\,\Psi_3 &= 2\,\frac{\Psi_1^3}{\rho^3} + \left( 9\,\alpha - \bar{\alpha} - \frac{3}{2} \frac{\overline{\Psi_1}}{\rho} \right) \frac{\Psi_1^2}{\rho^2} \\
&+ \left( 8\,\alpha^2 + 2\,\delta\,\alpha - 4\,\alpha \, \bar{\alpha} - 2\,\mu\,\rho - \Phi_1\overline{\Phi_1} - \frac{3\,\alpha\overline{\Psi_1}}{\rho} \right) \frac{\Psi_1}{\rho} + 2\,\rho\,\Psi_3.
\label{DPsi3}
}
\end{eqnarray}

\par
We now try to find an expression for $\delta\,\alpha$. In order to do so, we evaluate the $\delta$ derivative of equation (\ref{eq1_2}). Eliminating from the result $\delta\,\gamma$ and $\delta\,\bar{\gamma}$ by $NP_{15}$ and the complex conjugate of $NP_{18}$, we get
\begin{eqnarray}
\nonumber
\frac{1}{4}\left( 7\Psi_1^2 -9\Psi_1\overline{\Psi_1} +6\overline{\Psi_1}^2 \right) \frac{\Psi_1}{\rho^3}-{\frac{\overline{\Psi_1}^{3}}{{\rho}^{3}}} \\
\nonumber
+2\left( 3\left(2\Psi_1 - \overline{\Psi_1}\right)\alpha + \left( 2\overline{\Psi_1} - \Psi_1 \right) \bar{\alpha} \right) \frac{\Psi_1}{\rho^2}+\left( 3\,\alpha-11\,\bar{\alpha} \right) \frac{\overline{\Psi_1}^2}{\rho^2} \\
\nonumber
+\left( 4\,\delta\,\alpha +14\,\alpha^2 -2\,\alpha\,\bar{\alpha} -\Phi_1\overline{\Phi_1} \right) \frac{\Psi_1}{\rho}\\
\nonumber
-\left( 4\,\delta\,\alpha +4 \,\alpha^2 -14\alpha\,\bar{\alpha} +22\,\bar{\alpha}^2  -\Phi_1\,\overline{\Phi_1} \right) \frac{\overline{\Psi_1}}{\rho} \\
\nonumber
+ \left( \bar{\gamma}-\gamma -4\,\mu\right) \Psi_{{1}}+\left( 3\left(\bar{\gamma} -\gamma \right) +4\,\mu \right) \overline{\Psi_1} + 4\left(\nu-\bar{\nu} \right) \rho^2\\
\nonumber
+4 \left( \Delta\,\bar{\alpha} -\Delta\,\alpha + \left(  \gamma - \bar{\gamma} -2\,\mu \right)\alpha +\left( 3\left(\bar{\gamma} - \gamma \right) +2\,\mu \right) \bar{\alpha}   +\,\overline{\Psi_3} -\,\Psi_3 \right)\rho \\
+ 8\left( \left( \alpha - \bar{\alpha} \right) \delta\,\alpha  - \,\bar{\alpha}{\alpha}^{2} - \,\bar{\alpha}^{3} +2 \,\alpha\,\bar{\alpha}^{2}\right)
+ 4\left( \bar{\alpha}-\alpha\right) \Phi_1 \overline{\Phi_1}=0.
\label{delta_alpha}
\end{eqnarray}
Herewith the $[\delta, \Delta]\rho$-relation yields an expression for $\delta^2 \alpha$, while from the $[\bar{\delta}, \delta]\,\alpha$-relation, we get an expression for $\bar{\delta}\,\delta\,\alpha$. This then enables us to simplify $[\delta, \Delta]\,\Phi_1$, which results in an expression for $\Delta\,\alpha$ (note that the expressions we obtain all contain terms with factors $\delta\,\alpha$ and/or $\bar{\delta}\,\bar{\alpha}$ in the right hand side). Substituting the latter into (\ref{delta_alpha}), we eventually obtain :
\begin{eqnarray*}
\eqalign{
\rho^2 & \left(\Psi_1-\overline{\Psi_1}\right)\,\delta\,\alpha = -\frac{1}{16} \,\left(7\,\Psi_1^3 + \left(8\left(6\,\alpha-\bar{\alpha}\right)\rho -11\,\overline{\Psi_1}\right) \Psi_1^2 \right)\\
& +\frac{1}{16} \,\left( 16\,\mu\,\rho^3 -32\left(2\,\alpha-\bar{\alpha}\right)\alpha\,\rho^2 -4\left(5\,\bar{\alpha}-9\,\alpha\right)\rho\,\overline{\Psi_1} -7\,\overline{\Psi_1}^2\right)\Psi_1\\
& -\left( \overline{\Psi_3} -\Psi_3 \right) \rho^4 +\mu\,\overline{\Psi_1} \rho^3 +\left( -3 \,\bar{\alpha}^2\overline{\Psi_1} +2\,\alpha\,\bar{\alpha}\overline{\Psi_1} -\,\alpha^2\overline{\Psi_1} \right) \rho^2 \\
& -\frac{1}{16}\, \left( 12\, \alpha\,\overline{\Psi_1}^2 -36 \,\bar{\alpha}\,\overline{\Psi_1}^2 \right) \rho -3\, \overline{\Psi_1}^3.
}
\end{eqnarray*}
Now we can write $NP_{15}$ and $NP_{18}$ as expressions for $\delta\,\gamma$ and $\bar{\delta}\,\gamma$, after which we can solve (\ref{eq1_2}) for $\bar{\gamma}$ :
\begin{equation*}
\bar{\gamma} =\gamma+{\frac {\bar{\alpha}^{2}-{\alpha}^{2}}{\rho}}+{\frac {\bar{\alpha} \overline{\Psi_1}-\alpha\,\Psi_{{1}}}{{\rho}^{2}}}+\frac{1}{8} \,{\frac { \overline{\Psi_1}^{2}-{\Psi_{{1}}}^{2} }{{\rho}^{3}}}.
\label{gammabar}
\end{equation*}
Eliminating $\bar{\delta}\,\Psi_3$ from $[\bar{\delta}, \Delta]\,\Phi_1$, we can also solve $[\delta, \Delta]\,\Psi_1$ for $\Psi_4$ :
\begin{eqnarray*}
\eqalign{
\Psi_4 &= -\frac{1}{16}{\frac {{\Psi_{{1}}}^{4}}{{\rho}^{6}}} - {\frac {9}{8}} \left( \alpha - {\frac {1}{8}}\frac{\overline{\Psi_1}}{{\rho}} \right) \frac{{\Psi_{{1}}}^{3}}{\rho^5} - 3\left(   2\,\alpha^2 - \frac{1}{2} \frac{\alpha\,\overline{\Psi_1}}{\rho} + \frac{1}{32}\frac{\overline{\Psi_1}^2}{\rho^2} \right)\frac{\Psi_1^2}{\rho^4 }\\
&+\left( \frac{1}{2}\Psi_3 - 8\frac{\alpha^3}{\rho} + 3\frac{\alpha^2\overline{\Psi_1}}{\rho^2} -\frac{3}{8} \frac{\alpha\overline{\Psi_1}^2}{\rho^3} + \frac{1}{64} \frac{\overline{\Psi_1}^3}{\rho^4} \right) \frac{\Psi_1}{\rho^2}\\
&+\left(4\alpha - \frac{1}{2}\frac{\overline{\Psi_1}}{\rho} \right) \frac{\Psi_3}{\rho} .
\label{Psi4}
}
\end{eqnarray*}

\par
As in paper~\cite{Newman1}, we proceed by isolating the $\rho$-dependence of all variables. Equations (\ref{DPhi1}) and (\ref{DPsi1}) allow us to introduce variables $\phi_1$ and $\psi_1$, respectively defined by:
\begin{equation*}
\Phi_1 = \,\phi_1\,\rho, \qquad D\,\phi_1 = 0,
\end{equation*}
\begin{equation*}
\Psi_1 = \, \psi_1\,\rho^2, \qquad D\,\psi_1 = 0.
\end{equation*}
We also define a variable $L$ as
\begin{equation*}
L = \mathrm{log}\left| \rho \right|.
\end{equation*}
Since
\begin{equation*}
D\, (\alpha / \rho) = \frac{1}{4} \, \psi_1\,D\,L,
\end{equation*}
we see that
\begin{equation}
\label{alpha}
\alpha = \frac{1}{4}\,\psi_1\,L\,\rho+a\,\rho, \qquad D\,a = 0.
\end{equation}
This enables us to put $a$ equal to zero by a null rotation with parameter $\bar{B} =  \left(\frac{1}{4}\psi_1\,L\,\rho - \alpha \right)/\,\rho$ (as this is compatible with the condition $D\,B = 0$ (\ref{parameters})).
\par
Comparing the derivatives of $\alpha$ as in (\ref{alpha}) with those we already obtained before, we can extract expressions for $\lambda$, $\mu$ and $\nu$ :
\begin{equation*}
\lambda= \frac{1}{4}\left( L^2 + \frac{5}{2} L + \frac{3}{2} \right) \rho\,\psi_1^2 - \frac{1}{8} \left(3\,L + 1 \right)\rho\,\psi_1\overline{\psi_1} - \frac{1}{2}\,\rho\,L\,\phi_1\overline{\phi_1} - \frac{\Psi_3}{\rho\, \psi_1},
\end{equation*}
\begin{equation*}
\mu=\frac{1}{2} \left( \frac{1}{2} L^2 + \frac{7}{4} L + 1 \right) \rho\,\psi_1^2 - \frac{1}{4} \left( \frac{3}{2} L + 1 \right) \rho \psi_1\overline{\psi_1} - \frac{1}{2} \rho L\phi_1\overline{\phi_1} - \frac{\Psi_3}{\rho \psi_1},
\end{equation*}
\begin{eqnarray*}
\eqalign{
\nu &= \frac{1}{8} \left( \frac{1}{2} L^3 + L^2 - \frac{7}{4} L - \frac{5}{4} \right) \rho\psi_1^3 + \frac{1}{8} \left( \frac{1}{2} L^3 + \frac{1}{4} L^2 + L + 1 \right) \rho\psi_1^2\overline{\psi_1}^2  \\
&- \frac{1}{8} \Biggl[  \left( \frac{3}{4}\overline{\psi_1}^2 + \phi_1\overline{\phi_1} \right) L^2 + 2 \left( \frac{1}{8}\overline{\psi_1}^2 - \phi_1\overline{\phi_1} \right) L +\left( \frac{1}{4}\overline{\psi_1}^2 +\phi_1\overline{\phi_1}\right) \Biggr] \rho\psi_1\\
& + \frac{1}{2} \gamma\,\psi_1-\frac{1}{8} \left( L^2 + 1 \right) \rho\overline{\psi_1}\phi_1\overline{\phi_1}- \frac{1}{4} \left( \left( 1 + \frac{\overline{\psi_1}}{\psi_1} \right)  L - 2 \right)\frac{\Psi_3}{\rho}.
}
\end{eqnarray*}
Herewith we can integrate the $D\,\Psi_3$-equation (\ref{DPsi3}), which gives
\begin{equation*}
\Psi_{{3}}=\left( \frac{3}{16} \left( L + 3 \right) L \psi_1^3 -\frac{5}{16} L \psi_1^2\overline{\psi_1} - \frac{1}{2} L \psi_1\phi_1\overline{\phi_1}  + \psi_3 \right) \rho^2, \quad D\,\psi_3 = 0.
\label{Psi3}
\end{equation*}

\par
We now come to a more subtle part of the integration. First we introduce a help variable $\Omega = \delta\,\rho/D\,\rho$ and define two new operators $e_1$ and $e_2$ having the property that $e_1\,\rho = e_2\,\rho=0$ and $D\,e_1\,x = D\,e_2\,x=0$ for all $x$ obeying $D\,x=0$. One easily sees from the $[\delta, D]$-commutator that this can be achieved by putting $\delta =e_1 + \rho\,e_2 + \,\Omega\,D$ and $\bar{\delta} = \rho\, e_2 - e_1 + \bar{\Omega}\,D$. The new commutators $[e_1, D]$ and $[e_2, D]$ are then given by
\begin{equation*}
[e_1, D] = \left( [\delta , D] - [\bar{\delta}, D] \right)/2,
\end{equation*}
\begin{equation*}
[e_2, D] = \left( [\delta , D] + [\bar{\delta}, D] \right)/(2\rho) .
\end{equation*}
The expressions for the commutators $[e_1, \Delta] $ and $[e_2, \Delta] $ can be derived from $[\delta, \Delta] $ and $[\bar{\delta}, \Delta]$. They read
\begin{equation*}
[e_1, \Delta] = \frac{1}{16}\,\rho\,L \left( \overline{\psi_1}+\psi_{{1}}\right) ^{2}\,e_1,
\end{equation*}
\begin{eqnarray*}
\eqalign{
[e_2, \Delta] &= \frac{1}{4} \, \left( \psi_{{1}}^2 -\overline{\psi_1}^2  \right) e_1 +2\,\gamma\,e_2 -\frac{1}{8} \Biggl[ \left( L^2 + \frac{7}{2} L + 2 \right)\psi_1^2 \Biggr.\\
& + \left( \left( L + 3 \right)  \psi_1 - \frac{1}{2} \overline{\psi_1} \right) L \overline{\psi_1} \Biggl. + 2(L+1) \phi_1\overline{\phi_1} \Biggr] \rho e_2\\
& +\frac{1}{64} \Biggl[ \left( L+7 \right) \psi_1^3 - \left( 2L^2+ 5 L -3 \right) \psi_1^2\overline{\psi_1}\Biggr.  \\
&- \left( \left( 2 L^2+\frac{5}{4} L + \frac{3}{4} \right) \overline{\psi_1}^2 + 4\left( 2 L^2 + L - 1 \right) \phi_1\overline{\phi_1} \right) \psi_1 \\
&\Biggl. - 8 \left(  L^2 +2L -2\right) \overline{\psi_1}\phi_1\overline{\phi_1} + \left( L +1 \right) \overline{\psi_1}^3 - 32 \left( \frac{\overline{\psi_1}}{\psi_1} -  1 \right) \psi_3 \Biggr] D.
}
\end{eqnarray*}
Finally, from $[\bar{\delta},\delta]$ we get an expression for $[e_2, e_1]$ :
\begin{equation*}
[e_2, e_1] = -\frac{1}{4} \left( \psi_{{1}}+\overline{\psi_1} \right) e_1.
\end{equation*}
Note also that
\begin{equation}
\label{bare1}
\bar{e_1} = - e_1 \qquad \mathrm{and} \qquad \bar{e_2} = e_2.
\end{equation}
\par
At this stage we have expressions for all derivatives of $\gamma$, $\rho$ (thus of $L$), $\phi_1$, $\psi_1$ and $\psi_3$ ($\Delta \psi_3$ can be found from $NP_{10}$).

\par
Next we integrate the $D \, \gamma$-equation and obtain
\begin{equation*}
\gamma=g_{{0}}+\frac{1}{16} \left( \psi_{{1}}+\overline{\psi_1}  \right) \rho\,{L}^{2}\psi_{{1}}+\frac{1}{4} \,\rho L\,{\psi_{{1}}}^{2}+ \left( \frac{1}{8} \,{\psi_{{1}}}^{2}+\frac{1}{2} \,\phi_{{1}}\overline{\phi_1}  \right) \rho,
\label{gamma}
\end{equation*}
in which we can put $g_0$ equal to zero by a boost. To demonstrate this we first look at the derivatives of $\phi_1$, which are all zero, except for $\Delta\,\phi_1$, which equals $-4\,\phi_1\,\overline{\phi_1}\,g_0$. As a boost transforms $\phi_1$ into $\phi_1/A$, we can put $g_0$ to zero by choosing $\Delta\,\mathrm{log}A = -2\,g_0$. This also allows us to write $\phi_1$ as $Q\,f$, with $Q$ a non zero constant, and with $f$ on the unit circle ($\bar{f} = 1/f$).

\par
Now we write $\psi_1$ as $\psi_1 = (U + V)/2$ with $U$ real and $V$ imaginary. If we then look at the derivatives of $f$, $ \phi_1$, $ U$ and $ V$ we see that
\begin{equation*}
\mathrm{d}f = e_2\,f + \Delta\,f = \frac{f\,V}{8} \left( 4+ L \,\rho\,U\right) ,
\end{equation*}
\begin{equation*}
\mathrm{d} \phi_1 = e_2\,\phi_1 + \Delta\,\phi_1 = \frac{Q\,f\,V}{8} \left( 4 + L\,\rho\,U \right) ,
\end{equation*}
\begin{equation*}
\mathrm{d}U = e_2\,U + \Delta\,U = \frac{2\,V^2 - \, U^2 - 16\,Q^2 }{16} \left(4+ L\,\rho\,U \right),
\end{equation*}
\begin{equation}
\mathrm{d}V = e_2\,V + \Delta\,V = \frac{V\,U }{16} \left(  4+ L\,\rho\,U\right),
\label{e2V}
\end{equation}
showing that $f$, $\phi_1$, $U$ and $V$ are all functionally dependent. Notice that $U$ cannot be constant, as then $V$ would be real :
\begin{equation}
\label{e2U}
e_2\,U = \frac{V^2}{2} - \frac{U^2}{4} - 4\,Q^2.
\end{equation}
This permits us to use $U$ as a coordinate, but we prefer to write $U = U(x)$ (and hence also $V= V(x)$, $f = f(x)$ and $\phi_1 = \phi_1(x)$), with $x$ to be specified later.

\par
At this stage the only further information comes from introducing appropriate coordinates. First we construct the basis one forms $\Omega^1$, $\Omega^2$, $\Omega^3$, $\Omega^4$, dual to $e_1$, $e_2$, $\Delta$, $D$.
\par
The Cartan structure equations imply d$\Omega^3 = 0$ and thus $\Omega^3 = \mathrm{d}u$.
\par
Since $\mathrm{d}\rho = \Delta\,\rho\, \Omega^3 + 2\,\rho^2\, \Omega^4$ we have
\begin{equation*}
\Omega^4 = \frac{\mathrm{d}\rho}{2\,\rho^2} - \frac{\Delta\,\rho\,\mathrm{d} u}{2\,\rho^2}.
\end{equation*}
Next notice that $\mathrm{d}U = \Delta\,U\,\Omega^3 + e_2\,U\,\Omega^2$ and $\mathrm{d}V = \Delta\,V\,\Omega^3 + e_2\,V\,\Omega^2$, where, by (\ref{e2V}) and (\ref{e2U}), $e_2\,U$ and $ e_2\,V $ cannot be $0$. It follows that $\Omega^2$ can be written as a linear combination of $\mathrm{d} U $ and $\mathrm{d}u$. As $\Delta\,U = L\,\rho\,U \left( 2\,V^2-U^2 - 16\,Q^2 \right) /16 $, we can write
\begin{equation*}
\Omega^2 = S(x)\,\mathrm{d}x - \frac{U\,L\,\rho}{4}\mathrm{d}u,
\end{equation*}
with $S(x)$ to be specified below. There remains $\Omega^1$ which is general, $\Omega^1 = B\,\mathrm{d}u + C\,\mathrm{d}\rho + H\,\mathrm{d}x + J \,\mathrm{d}y$, but where we can choose $C = 0$ by a transformation of the $y$-coordinate. From (\ref{bare1}) we see that
\begin{equation*}\begin{array}{lll}
\bar{J} &=& -J,\\
\bar{S} &=& S,\\
\bar{H} &=& -H.
\end{array}
\end{equation*}
The tetrad basis vectors are then given by
\begin{equation}
\label{e1}
e_1=\frac{1}{J}\frac{\partial}{\partial y},
\end{equation}
\begin{equation}
\label{e2}
e_2=\frac{1}{S}\frac{\partial}{\partial x}-\frac{H}{S\,J}\frac{\partial}{\partial y},
\end{equation}
\begin{equation}
\Delta=\frac{\partial}{\partial u}+ \Delta\,\rho \frac{\partial}{\partial \rho} + \frac{L\,\rho\,U}{4\,S}\frac{\partial}{\partial x} - \frac{4\,B\,S+H\,L\,\rho\,U}{4\,S\,J}\frac{\partial}{\partial y},
\end{equation}
\begin{equation}
\label{D}
D=2\,\rho^2 \frac{\partial}{\partial \rho}.
\end{equation}
Acting now with the commutators on the coordinates $u$, $\rho$, $x$ or $y$, we get
\begin{eqnarray*}
D\,J = D\,S = D\,H = 0 = e_1\,S , \\
\Delta \,S =\frac{1}{4} \,L\rho\,U\,e_2\,S, \\
D\,B = \frac{1}{2}\,\rho\,V.
\end{eqnarray*}
It can then easily be seen that
\begin{equation*}
B=\frac{1}{4} \,L\,V+ B_0,\qquad D\,B_0 = 0.
\end{equation*}
From $D\, B_0 = 0 = D\,H = D\,J$ one deduces the existence of a new $y$-coordinate, as a function of $u$, $x$ and $y$, such that $H$ becomes $0$.

\par
From $[e_2,e_1]y$ we find the following expression for $e_2\,J$ :
\begin{equation*}
	e_2\, J= \frac{JU}{4} \quad \mathrm{or} \quad \frac{\mathrm{d}\,\mathrm{log}J}{\mathrm{d} x} = \frac{S\,U}{4},
\end{equation*}
from which we see that $J = J_0(x)J_1(u,y)$. Defining a new $y$ coordinate as a function of $u$ and $y$ and absorbing the $\mathrm{d}u$-part in a new $B$-coefficient, one can assume $J_1 =1$. Herewith $J = J(x)$ and we have that $e_1\,J=0$ and $\Delta \,J = L\,\rho\,U^2 J /16$.

\par
From $[e_1, \Delta] y$ we find that $e_1 \, B_0 = 0$, which means that we can write $B_0 = J\, B_1(x,u)$, after which $[e_2,\Delta] y$ results in \begin{equation}\label{B1vgl} e_2 \, B_1 = -UV/(4J).\end{equation} Herewith $B_1$ can be decomposed as $B_1 = B_2 (x) + B_3 (u)$. A final $u$-dependent $y$-translation allows to transform $\Omega_1$ into $\Omega_1 = \left( VL/4 + JB_2(x)\right) \mathrm{d} u + J \mathrm{d}y$, $i.e.$ one can assume $B_1 = B_1(x)$.

\par
We now fix $S(x)$ such that we can integrate (\ref{B1vgl}) : As $e_2\,V = UV/4$ and $e_2\,J = UJ/4$, the choice 
\begin{equation}
S(x)= 4/(x\,U) 
\end{equation}
leads to
\begin{equation*}
	B_1 = -\frac{V}{J} \mathrm{log}x.
\end{equation*}

From the $e_2\,V$-equation, we see that $V = 4\,Q\,a\,x$, with $Q\,a$ constant ($Q$ is assumed to be non zero).
We still need a solution for $U$. From the expressions for $D\,U$, $\Delta\,U$, $e_1\,U$ and $e_2\,U$ it is easy to see that $U=U(x)$ with $U(x)$ determined by
\begin{equation}
\frac{\partial \, U}{\partial\,x} = -{\frac {16\,{Q}^{2}+32\,{Q}^{2}{a}^{2}{x}^{2}+{U}^{2}}{xU}},
\label{ddxu}
\end{equation}
which integrates to
\begin{equation}
{x}^{2}{U}^{2}=16 \left( {C}^{2}-Q^2 {a}^{2}{x}^{4}-Q^2 {x}^{2} \right).
\label{equ}
\end{equation}
From $e_1$, $e_2$, $\Delta$ and $D$ applied to $\left( \psi_3 + \overline{\psi_3}\right)/2$ we see that the partial derivatives of $\psi$ (the real part of $\psi_3$ divided by $U$) are given by :
\begin{equation*}
	\frac{\partial \psi}{\partial y} = 0 =\frac{\partial \psi}{\partial u} =\frac{\partial \psi}{\partial \rho},
\end{equation*}
\begin{equation*}
\label{dpsidx}
\frac{\partial \psi}{\partial x} = \frac{U^2 + 16\,(1 - 14\, x^2 a^2) Q^2 -64 \,\psi}{32\, x}.
\end{equation*}
Using the equation (\ref{equ}) we can solve these differential equations for $\psi$ :
\begin{equation*}
	\psi= \frac{1}{8} \left( 4\, c_1^2\mathrm{log}x + 8 \,c_2 - 15\, a^2x^4 \right) \frac{Q^2}{x^2}, \quad \mathrm{with} \quad c_1,c_2 \quad \mathrm{constant}.
\end{equation*}
\par
The dual basis now becomes
\begin{equation*}
\Omega^1 = \left( \frac{1}{8} \left( 4 I Q a x-U \right) L -2\,I Q a x \mathrm{log}x\right) \mathrm{d}u+ {\frac {2}{xU\rho}}\mathrm{d}x+\frac{1}{2}I x J_0\mathrm{d}y,
\end{equation*}
\begin{equation*}
\Omega^2 =\left( -\frac{1}{8} \left( 4 I Qax + U \right) L + 2\,IQax\mathrm{log}x \right)\mathrm{d}u + {\frac {2}{xU\rho}}\mathrm{d}x-\frac{1}{2} IxJ_0 \mathrm{d}y,
\end{equation*}
\begin{equation*}
\Omega^3 =\mathrm{d}u,
\end{equation*}
\begin{eqnarray*}
\eqalign{
\Omega^4 &= \frac{1}{2\,\rho^2}\mathrm{d}\rho-\frac{1+L}{2\,x\,\rho}\mathrm{d}x-\frac{1+L }{2}J_0 Q x^2a\mathrm{d}y+ \left[ \frac{1}{64} \left( 2+L\right) LU^2 - 2\,\psi + \right.\\
&\left. \left( 2 \left( \mathrm{log}x - 2 \right) x^2a^2 + \frac{1}{2} \left( 1 + \left( 4 \mathrm{log}x-1\right) x^2a^2\right) L - \frac{3}{4} x^2a^2L^2 \right)Q^2 \right]\mathrm{d}u,
}
\end{eqnarray*}
which shows that we can absorb $J_0$ in $y$.

\par
The metric becomes
\begin{eqnarray}
\eqalign{
\mathrm{d}s^2 &= \left(\left(2 L - 4  \mathrm{log}x + 1\right)Q x^2a \mathrm{d}y - \frac{1}{\rho^2}\mathrm{d}\rho + \frac{2}{\rho x} \mathrm{d}x \right) \mathrm{d}u  +\frac{1}{2} \,{x}^{2}{\mathrm{d}y}^{2}\\
&+ \left[ 2L^2a^2x^2 - \left( \frac{c_1^2}{x^2} + (8\mathrm{log}x - 2) a^2x^2 \right) L + 2 \left( \frac{c_1^2}{x^2} - 2 a^2x^2 \right)\mathrm{log}x \right.\\
&\left. + \frac{4 \, c_2}{x^2} + \frac{a^2 x^2}{2} \right] Q^2 \mathrm{d} u^2 + \frac{1}{2\left( c_1^2 - x^2- a^2 x^4 \right) \rho^2Q^2 } \mathrm{d}x^2,
\label{lijnel}
}
\end{eqnarray}
with the Maxwell field given by
\begin{equation}
\label{P0}
\Phi_0 = 0,
\end{equation}
\begin{equation}
\label{P1}
\Phi_1 = Q f\rho,
\end{equation}
\begin{equation}
\label{P2}
\Phi_2 =\frac{1}{4}\left(LU + 4(L+1)IQax\right)Qf\rho,
\end{equation}
where
\begin{equation*}
f=-\frac{2\,a\sqrt{c_1^2-(a^2x^2+1)x^2} + (2\,a^2x^2 + 1) I}{\sqrt{4\,a^2c_1^2+1}}
\end{equation*}
is on the unit circle, as required.

A more elegant expression for the metric (\ref{lijnel}), preserving
the obvious symmetries of the solution, is obtained by a coordinate transformation $\rho \rightarrow r\, x^2 $, which results in
\begin{eqnarray*}
\eqalign{
ds^2 &=\frac{1}{2(c_1^2-x^2-a^2x^4)r^2x^4Q^2}\mathrm{d}x^2 + \left( a(2L+1)Qx^2\mathrm{d}y - \frac{1}{r^2 x^2} \mathrm{d}r \right)\mathrm{d}u\\
&+\left( \frac{a^2(2L+1)^2Q^2x^2}{2}-\frac{(Lc_1^2-4c_2)Q^2}{x^2}\right) \mathrm{d}u^2 + \frac{x^2}{2}\mathrm{d}y^2,
}
\end{eqnarray*}
in which $L = \mathrm{log}r = \mathrm{log}\rho-  2\,\mathrm{log}x$.

\section{Vacuum limit}

Neither $u$ nor $y$ appear in the components of the Riemann tensor $\Psi_1$, $\Psi_2$, $\Psi_3$, $\Psi_4$ and $\Phi_0$, $\Phi_1$, $\Phi_2$ (\ref{P0})-(\ref{P2}):
\begin{equation*}
\Psi_1 =\frac{\left(U + 4\,IQax\right)\rho^2}{2},
\end{equation*}
\begin{equation*}
\Psi_2 =\frac{\left(U + 4\,IQax\right)\rho^2}{8}\left((L+1)U+4(L+2)IQax\right),
\end{equation*}
\begin{eqnarray*}
\eqalign{
\Psi_3 &=\frac{ \left(U+4\,IQax\right)\rho^2}{128 U}\Bigl[ (3L+4)U^3L + 24(L^2+3L+1)IQxaU^2 \Bigr.\\
&\Bigl. -16 \left( 2L + (3L^2+14L+9)x^2a^2-\frac{4\mathrm{log}x c_1^2+8c_2}{x^2}\right)Q^2U\Bigr],
}
\end{eqnarray*}
\begin{eqnarray*}
\eqalign{
\Psi_4 &= \frac{ \left( U+4IQax\right)\rho^2 }{256 U}\left(LU+4\left(L+1\right)IQax\right)\biggl[(8L^2+32L+12)IQxaU^2  \biggr.\\
&\biggl. +(L+1)U^3L -16 \left(2L + (L^2+7L+4)x^2a^2- \frac{4\mathrm{log}xc_1^2+8 c_2}{x^2} \right)Q^2U\biggr],
}
\end{eqnarray*}
and one can easily see that also in the canonical Petrov type I tetrad precisely two functionally independent functions will remain in the curvature components. Therefore the metric will admit exactly two Killing vectors~\cite{karlhede}. The vacuum limit of our metric then can not be the general one obtained by Newman and Tamburino, as the latter has only one Killing vector.
In fact the ($Q=0$)-limit of (\ref{lijnel}) is given by
\begin{equation*}
{\frac {(2\mathrm{log}x-L)c_0^2}{{x}^{2}}}{\mathrm{d}u}^{2} + \left( 2\,{\frac {\mathrm{d}x}{x\rho}}-{\frac {\mathrm{d}\rho}{{\rho}^{2}}} \right) \mathrm{d}u + \frac{1}{2} \,{\frac {{\mathrm{d}x}^{2}}{{c_0}^{2}{\rho}^{2}}}+\frac{{x}^{2}}{2}{\mathrm{d}y}^{2},
\end{equation*}
which, after the coordinate transformations $x \rightarrow c_0\,x/ \sqrt{2}$, $\rho \rightarrow -1/(2\,r)$, $y \rightarrow 2\,y / c_0$, equals (26.23) of~\cite{Kramer}.

\section{Conclusion}
We have obtained the general solution of the `aligned Newman-Tamburino-Maxwell' problem. These are the algebraically general metrics possessing hypersurface orthogonal geodesic rays, with non vanishing shear and divergence, in the presence of an \emph{aligned} Maxwell field. We have shown that the solution necessarily satisfies the cylindrical condition $|\rho|=|\sigma|$ and has exactly two Killing vectors, which implies that the vacuum limit is only a subfamily of the general cylindrical Newman Tamburino metrics.

\section*{Acknowledgment}
 NVdB expresses his thanks to Robert Debever\footnote{(1915-1998)}
 and Jules Leroy (Universit\'e Libre de Bruxelles), whose questions about the Newman Tamburino metrics have led to the present investigation.

\section*{References}

\providecommand{\newblock}{}

\end{document}